\documentclass[a4paper,12pt]{article}
\usepackage{graphicx}
\usepackage{amssymb}
\usepackage{amsmath}
\usepackage[a4paper,portrait,twocolumn=false,includeheadfoot,mag=1000,dvips]{geometry}
\usepackage[title]{appendix}
\usepackage{caption}
\usepackage{float}
\usepackage{color}
\usepackage{ulem}
\usepackage{mathrsfs}
\usepackage{type1cm}
\usepackage{marvosym}
\usepackage{amssymb}
\usepackage{wasysym}
\usepackage{subfigure}
\usepackage{multirow}
\usepackage{booktabs}
\usepackage[utf8]{inputenc}

\usepackage{graphicx}
\usepackage{epstopdf}

\usepackage{makecell}

\usepackage{enumerate}

\makeatletter

\newcommand{\Rmnum}[1]{\expandafter\@slowromancap\romannumeral #1@}
\makeatother

\numberwithin{equation}{section}

\newcommand{\xiaosihao}{\fontsize{14pt}{\baselineskip}\selectfont}

\begin{document}  \xiaosihao

\newtheorem{The}{Theorem}
\newtheorem{exam}{Example}
\newtheorem{lem}{Lemma}
\newtheorem{de}{Definition}
\newtheorem{prop}{Proposition}
\newtheorem{cor}{Corollary}
\newtheorem{hyp}{Hypothesis}
\newtheorem{rem}{Remark}
\pagestyle{plain}

\setlength{\baselineskip}{20pt}
\setlength{\parskip}{0.4\baselineskip}

\clearpage
%
%
%
%
%
%
%

%

\title{What Does the ``Mean'' Really Mean? }
\author{\normalsize by \\\relax Nozer D. Singpurwalla \\\relax  The George Washington University, Washington, D.C., USA and \\\relax Boya
	Lai  \\\relax The City University of Hong Kong, Hong Kong}
\maketitle

\begin{abstract}
{\normalsize
	The arithmetic average of a collection of observed values of a homogeneous collection of quantities is often taken to be the most representative observation. There are several arguments supporting this choice the moment of inertia being the most familiar. But what does this mean?\\ [-15pt]
		
	In this note, we bring forth the Kolmogorov-Nagumo point of view that the arithmetic average is a special case of a sequence of functions of a special kind, the quadratic and the geometric means being some of the other cases. The median fails to belong to this class of functions. The Kolmogorov-Nagumo interpretation is the most defensible and the most definitive one for the arithmetic average, but its essence boils down to the fact that this average is merely an abstraction which has meaning only within its mathematical set-up.\\ [10pt]}
{\normalsize \textbf{Keywords:} \textit{Chisini's Equation, Kolmogorov-Nagumo Functions, Weighted Means}.\\}
\end{abstract}
\addtocounter{section}{-1}

\section{\large Background}
The December 2017 issue of ``\textit{\textbf{Significance}}'', an ASA co-sponsored magazine, published an engaging article by Simon Raper titled, ``The Shock of the Mean''. A title like this may come as a surprise to today's statisticians because most are not shocked when they encounter a mean, taken here to be an arithmetic average. According to Raper, the 18-th century shock had to do with how the mean was used, and what it meant. It had little to do with the mathematical underpinnings of the mean because these became transparent only during the 1930's. The purpose of this article is to articulate on these underpinnings which go beyond the usual explanations, like the mean is a moment of inertia.  The mean continues to be an abstraction with an interpretation only within its mathematical framework; it may therefore continue to shock many a modern statistician who has wholeheartedly embraced it. But first some words about the merits of Raper's article.

Fundamentally, Raper's article is of expository value. It gives a fascinating discourse on the notion of an arithmetic mean by tracing its historic roots, providing anecdotal stories connected with its appearance and its acceptance, and its evolution as a commonly used methodological device in the economic, the engineering, the medical, the and the social sciences. Of note on p. 15, is a timeline of the mean starting from 426 BC until 1810, when Laplace published his central limit theorem. The material to be given in this entry has a timeline subsequent to 1810, and its focus is more on the analytics of the mean.

At about the same time as the appearance of the Raper article, the authors of this entry were looking at Shannon's formulas for entropy and information. A central feature of these formulas is a weighted average of the ``information gained'' in every realization of a random variable. In the course of appreciating the essence of this operation, the authors encountered two papers, one by George Barnard (1951) and the other by Alfred Renyi (1961).  Both papers questioned Shannon's rationale for choosing the weighted average, something which seemed like a natural thing to do. In this context, Renyi also mentioned (without any reference) the \textbf{\textit{Kolmogorov-Nagumo}} class of functions, of which the sample average turns out to be a special case. Indeed, Renyi used this class of functions (involving improper random variables), to propose his measure of information. All of this seemed intriguing, and on pursuing the matter further, it became clear that outside the community of functional analysts, little has been said about this class of functions, a special case of which is a statistician's most basic tool. Even Stigler's (2016) masterpiece \textit{\textbf{The Seven Pillars of Statistical Wisdom}} does not seem to make note of this foundation on which one of his pillars rests. In what follows, we highlight the mathematical essence of the mean which has a history dating back to the times of Cauchy.

\section{\large Antecedents to the Kolmogorov-Nagumo Functions}

The earliest reference to the mathematical notion of a \textit{\textbf{mean}}, is that it is  \textit{\textbf{a class of functions}}, say M, of n measurements $x_1,\cdots,x_n$,  on a homogeneous collection of n quantities satisfying a certain condition. It is due to Cauchy (1821). All that Cauchy required is that $M(x_1,\cdots,x_n)$, be bounded by the smallest and the largest values of $x_1,\cdots,x_n$, as:
$$min\{x_1,\cdots,x_n\}\leq M\{x_1,\cdots,x_n\}\leq max\{x_1,\cdots,x_n\}$$

Whereas Cauchy does not give an interpretive meaning to the function M, he initiated a pathway for much that followed, leading up to the definitive works of Kolmogorov (1930) and of Nagumo (1930). However, the notion that the value taken by certain members of the class of functions M may be seen as a representative measurement of the measurements $x_1,\cdots,x_n$, is ascribed to Chisini (1929); see Marichal (2000). Chisini was a distinguished Italian geometer, who was de Finetti's teacher at the University of Milan.

Subsequent to Cauchy (1821), but prior to Chisini (1929), is an exhaustive paper, with discussion, by John Venn (1891) titled `` \textit{\textbf{On the Nature and Uses of Averages}}'', that he read before the Royal Statistical Society. Whereas Cauchy's perspective is analytical, Venn's has more to do with applications of the arithmetic average. Specifically, Venn raises several questions related to the average. He asks: ``Why resort to averages at \mbox{all}''? ``What do we gain and lose respectively, by doing so''? What different kinds of averages are there, and how and why does one such kind become more appropriate than another''? Venn, via a footnote, also states that a mathematical justification of almost every kind of average can be found in Edgeworth's paper in the Cambridge Philosophical Transactions. Whereas we have not been successful in accessing Edgeworth's paper, it appears that Chisini, if not Bonferroni (1927), may have come close to answering many of the questions raised by Venn.

Per Chisini, a  \textit{\textbf{representative value}} of $x_1,\cdots,x_n$, with respect to the function M, is a number $\mu$ such that if each of the $x_i$'s are replaced by $\mu$, the value of the function M is unchanged. That is:
$$M(\mu,\cdots,\mu)=M(x_1,\cdots,x_n)$$
this is known as \textit{\textbf{Chisini's Equation}}.

When the function M is the sum of its arguments, the solution to the Chisini Equation is the arithmetic mean, known to statisticians as the \textit{\textbf{sample mean}}. Similarly, when M is the product (the sum of squares) [the sum of inverses] {the sum of exponentials}, then the solution to this equation is the geometric mean (the quadratic mean) [the harmonic mean] {the exponential mean}.

As an illustration, suppose that M is additive, so that $M(\mu,\cdots,\mu)= n \mu = M(\sum x_i)$; then $\mu =\Sigma x_i /n$, the \textit{\textbf{arithmetic mean}}. Similarly, if the function M connotes a product, so that  $M(\mu,\cdots,\mu)=\mu^n = M(\prod_{i=1}^n x_i)$, then $\mu$ is the \textit{\textbf{geometric mean}}. With M as the sum of squares, $\mu = \sqrt{\frac{1}{n}\sum x_i^2}$, is the \textit{\textbf{quadratic mean}}, whereas with M as the sum of inverses, $\mu = \frac{1}{\frac{1}{n}\sum \frac{1}{x_i}}$, the \textit{\textbf{harmonic mean}}.

To summarize, the commonly used measures of representative values, referred to as measures of central tendency, are effectively, solutions to Chisini's equation. Preceding Chisini, is the work of Bonferroni (1924), who after Cauchy may have set the stage for that which is to follow [cf. Muliere and Parmigiani (1993)].

The story would end here with a statement about the solution to Chisini's equation, except for a caveat. This has to do with the fact that a solution to the equation, assuming it exists, may not satisfy Cauchy's inequality; this fact has been pointed out by de Finetti (1931). Indeed, de Finetti's motivation in writing this paper was more ambitious. He wanted to extend Chisini's definition of the mean of a collection of measurements, to that of the mean of a collection of functions, particularly, probability distribution functions [see Cifarelli and Regazzini (1996)]. More important, de Finetti was endeavouring to connect the notion of the mean with the notion of a \textbf{\textit{certainty equivalent}} in decision and utility theory [see Muliere and Parmigiani (1993)].

Recognizing that the notion of a mean should be more than a function M which merely satisfies Cauchy's condition, or which is a solution to Chisini's equation, Kolmogorov and Nagumo, independently and simultaneously, proved a fundamental theorem about mean values.

\section{\large The Kolmogorov-Nagumo Theorem on Means.}

Kolmogorov (1930), and Nagumo (1930), henceforth K-N, respectively, propose a definition of a mean in terms of a sequence of a family of functions, and provide a theorem to operationalize them.  Specifically, the mean is an infinite sequence of functions $M_1 (x_1 ),M_2 (x_1,x_2 ),M_3 (x_1,x_2,x_3 ),\cdots,M_n (x_1,x_2,\cdots,x_n)$, each $M_n$ being continuous, increasing, and symmetric, and with the property that $M_n (x,x,\cdots,x)=x$, for all x, and all n; a reflexive law. Furthermore, the terms of this sequence are related by an associative law of the following nature:
$$M_k (x_1,x_2,\cdots,x_k)=x\Rightarrow M_n (x_1,\cdots,x_k,x_{k+1},\cdots,x_n)= M_n (x,\cdots,x,x_{k+1},\cdots,x_n),$$
for every integer $k\leq n$.

The striking theorem of K-N [cf. Aczel (1948)], is that under the above necessary and sufficient conditions on the above sequence of function (also known as the Kolmogorov-Nagumo funtions), there exists a continuous and strictly increasing function $f$ by which the mean value $M_n (x_1,\cdots,x_n )$ can be written as:
$$M_n (x_1,x_2,\cdots,x_n) = f^{-1}[\frac{1}{n}\sum_{1}^{n}f(x_i)],$$
where $f^{-1} (x)$ is the inverse of $f(x)$.

Different choices for $f(x)$ yield different functional forms for the mean $M_n$. For example, if $f(x)=x$, then $M_n (x_1,x_2,\cdots,x_n)=\frac{1}{n}\sum x_i $ , the arithmetic mean. Similarly, if $f(x)=x^2$, then the mean is the quadratic mean. The table below gives a summary of some choices for $f$.

\begin{table}[htbp]
	\centering
	\begin{tabular}{m{0pt}p{60pt}<{\centering}p{60pt}<{\centering}p{60pt}<{\centering}}
		\hline
		\rule{0pt}{18pt} & Choice of $f(x)$   & Mean $M_n(x_1,\cdots,x_n)$ & Qualifier of Mean       \\
		\hline
		\rule{0pt}{20pt} &$x$   & $\frac{1}{n}\sum n_i$ & Arithmetic      \\
		\rule{0pt}{20pt} &$x^2$   & $\sqrt{\frac{1}{n}\sum n_i^2}$ & Quadratic      \\
		\rule{0pt}{20pt} &$\log x$   & $\sqrt[n]{\prod x_i}$ & Geometric      \\
		\rule{0pt}{20pt} &	$\frac{1}{x}$   & $\frac{1}{\frac{1}{n}\sum\frac{1}{x_i}}$ & Harmonic      \\
		\rule{0pt}{20pt} &$x^{\alpha}$   & $(\frac{1}{n}\sum x_i^{\alpha})^{\alpha}$ & Power      \\
		\hline
	\end{tabular}
\end{table}

The fact that the median of n measurements $x_1,\cdots,x_n$, does not belong as an entry in the table above, was remarked by de Finetti (1931). This is because the median does not obey the associative law. Thus per the K-N criteria, the median cannot be seen as a representative measurement of the n measurements. In the same 1931 paper, de Finetti, and also Kitagawa (1934) generalized the K-N result in the case of weighted observations. If for any observation $x_i$ there is associated a weight $q_i$, with $\sum q_i =1$, then de Finetti and Kitagawa gave conditions for writing
$$M_n (x_1,\cdots,x_n;q_1,\cdots,q_n) = f^{-1}[\sum q_if(x_i)],$$
for $n=1,2,\cdots$. Weighted means are germane in contexts like Bayesian decision making wherein taking expected utilities is a necessary step, and each $x_i$ is associated with a utility.

The only justification for taking expected values we know of is in decision theory which envolves choosing that decision which maximizes an expected utility.

\section{\large Concluding Remarks.}
The answer to the question ``What Does the ``Mean'' Really Mean'' posed in the title has gone from the very verbal and descriptive like ``representative measurement'', to the physical like ``first moment'', to the mathematical and abstract like ``the Kolmogorov-Nagumo sequence of functions''. The Kolmogorov-Nagumo focused answer seems most definitive and final, though it suffers from the fact that neither Kolmogorov nor Nagumo say much, if anything, as to what the function $f$ should be. Rather, theirs is a statement about the existence of $f$ and about and exclusion, like the median. Precursors to the Kolmogorov-Nagumo work see the mean as merely a function per Cauchy, or the solution to an equation per Chisini.

\section*{ Acknowledgements}

Fabrizio Ruggeri was the cause of making the authors aware of the exhaustive and thorough paper of Muliere and Parmigiani. Thanks Fabrizio.

\end{document}